# Topological Interface-State Lasing in a Polymer-Cholesteric Liquid Crystal Superlattice

Yu Wang,[1,*] Donghao Yang,[1,*] Shaohua Gao,[2,*] Xinzheng Zhang,[1,4‡] Irena Drevensek-Olenik,[3]

Qiang Wu[1], Marouen Chemingui,[1] Zhigang Chen[1,4‡], and Jingjun Xu[1‡]

*1 The MOE Key Laboratory of Weak-Light Nonlinear Photonics and International Sino-Slovenian Joint Research Center on Liquid Crystal Photonics, TEDA Institute of Applied Physics and School of Physics, Nankai University, Tianjin 300457, China*

*2 Institute of Optoelectronic Engineering, College of Physics & Optoelectronics, Taiyuan University of Technology, Taiyuan 030024, China*

*3 Faculty of Mathematics and Physics, University of Ljubljana, and Department of Complex Matter, J. Stefan Institute, SI-1000 Ljubljana, Slovenia*

*4 Collaborative Innovation Center of Extreme Optics, Shanxi University, Taiyuan, Shanxi 030006, China*

[*]*These authors are co-first authors of the article.*

‡ *zxz@nankai.edu.cn; zgchen@nankai.edu.cn; jjxu@nankai.edu.cn*

**Abstract:** The advance of topological photonics has heralded a revolution for manipulating light as well as for the development of novel photonic devices such as topological insulator lasers. Here, we demonstrate topological lasing of circular polarization in a polymer-cholesteric liquid crystal (P-CLC) superlattice, tunable in the visible wavelength regime. By use of the femtosecond-laser direct-writing and self-assembling techniques, we establish the P-CLC superlattice with a controlled mini-band structure and a topological interface defect, thereby achieving a low threshold for robust topological lasing at about 0.4 μJ. Thanks to the chiral liquid crystal, not only the emission wavelength is thermally tuned, but the circularly polarized lasing is readily achieved. Our results bring about the possibility to realize compact and integrated topological photonic devices at low cost, as well as to engineer an ideal platform for exploring topological physics that involves light-matter interaction in soft-matter environments.

## Introduction

The emergence of topological physics [1] has brought a new degree of freedom into photonics and, consequently, opened up a new era of topological photonics [2-5]. Following conventional bulk-edge correspondence principle, the *N*-dimensional topological insulators can host *N*-1-dimensional topological edge states [3]. These states are topologically protected and hence are robust and immune to scattering from defects, impurities, and other kinds of disorder, as demonstrated with electromagnetic waves [6-8]. Along with recently discovered higher-order topological insulators [9-11], nontrivial topological phases can be utilized to implement a variety of topological photonic devices such as broadband unidirectional propagating waveguides, optical isolators, optical splitters and robust optical delay lines [12-16], and, in particular, the topological resonators and lasers [17-25]. Robust lasing has been demonstrated also by employing one-dimensional chiral edge states and two-dimensional valley-Hall edge states in various photonic settings [26-34], or by use of the topological features of non-Hermitian exceptional points [35]. Overall, it has been proven that optical modes with topological protection can significantly improve the laser performance. At present, most of the topological lasers are based on photonic structures of characteristic size typically on the micrometer scale, with lasing wavelengths at the infrared wavelength region. On the other hand, laser sources in the visible spectral region play an important role in for example chemical detection, biology, and medicine [36,37], so it is imperative to develop visible topological lasers. Recently, a planar topological laser based on a topological structure with a thin layer of an active organic material operating in the red spectral region has been reported [38]. Despite those efforts, thus far it still remains a challenge to fabricate topological structures for lasing in visible spectral region with high performance, especially when tunability and controlled polarization characteristics are also desired. This brings about a question: is it possible to employ natural materials exhibiting photonic bandgaps to realize a low-cost tunable topological laser to overcome such a challenge?

In this work, we establish a photonic structure consisting of an assembly of laser active polymer ribbons forming a topological Su-Schrieffer-Heeger (SSH) lattice [39-46] filled with a cholesteric liquid crystal (CLC). Two such polymer CLC (P-CLC) lattices are fabricated to contain a short-short defect (SSD) at the interface between topologically trivial and nontrivial regimes [45-47], whereas the CLC serves as a self-assembled optical bandgap medium to achieve desired tunability. With such a composite photonic structure, we demonstrate topological interface-state (TIS) lasing with circular polarization at visible wavelengths, and compare its lasing threshold with trivial defect state lasing. The presence of polymer ribbons enables the formation of mini-bands in the inherent photonic bandgap from the CLC, whereas the topological properties are manipulated by appropriately arranging intra- and inter-cell coupling strengths in the SSH lattices. Furthermore, we demonstrate thermal tuning of the lasing wavelengths, and show that the topological lasing from such superlattices is robust against chiral perturbations. While liquid crystals have been widely explored for applications in lasers [48-50] and multi-color reflectors [51], we realize here for the first time to our knowledge low-threshold topological lasing of circular polarization based on CLCs.

## Results

### Topological P-CLC superlattices and emission spectrum

Liquid crystals are an important area in the field of soft matter photonics, and it is especially attractive to break the inversion symmetry by utilizing chiral materials such as CLCs [52-54]. Our assembly belongs to the category of one-dimensional (1D) coupled arrays forming the well-celebrated SSH lattice [39,40], which incorporates a CLC possessing a spontaneously-formed periodic helical structure. Such CLC structures exhibit a photonic bandgap for circularly polarized light, with helicity matching the CLC. The spectral properties of this photonic bandgap can easily be modulated by external stimuli, such as an electric or magnetic field, pressure, light, and temperature. Therefore, tunable lasers can be readily realized based on CLCs [55, 56]. Here we employ such unique properties of CLCs to realize tunable topological lasers at low cost. We analyze the coupling of defect modes in the P-CLC superlattices and demonstrate that the mini-bands can be designed in the bandgap of the CLC, which has been proposed to achieve visible defect mode lasing with a low threshold [57]. In experiments, we apply an out-of-plane liquid crystal orientation technology based on the femtosecond laser direct writing (FLDW) method recently developed by our group [58]. The corresponding P-CLC superlattices benefit from a combination of optical activity and tunability characteristics of the CLCs, stability and processability characteristics of the polymer materials, and unique features of the topological structure.

Figure 1(a) shows a schematic drawing of the corresponding topological structure and the intensity profile of the associated TIS. Specifically, the aforementioned SSD topological structure consists of parallelly oriented polymer ribbons doped with a fluorescent dye (PM597). The space between the ribbons is filled with a right-handed CLC (RHCLC) (see Materials and Methods). Surface relief gratings present on the sidewalls of the ribbons ensure that the CLC is well aligned (out-of-plane orientation technology) [59]. Its helical axis is perpendicular to the ribbons. The helical periodicity $p$ (pitch) of the CLC, which is determined by the concentration of the doped chiral agent and the temperature, is assumed to be $p = 348$ nm here. As shown in Fig 1(a, b1), the distances between the weakly and strongly coupled ribbons are chosen to be $d_1 = 34 \cdot (p/2)$ and $d_2 = 22 \cdot (p/2)$, respectively. The width of each polymer ribbon is set to be $d_i = 2$ μm, which is thick enough to give rise to two defect modes in the bandgap of the CLC [57]. The two P-CLC SSH superlattices in either side of the interface have the same mini-band structure, but they possess different topological properties that are characterized by their winding numbers [60-63]. A topological phase transition occurs through the band inversion, resulting in two non-trivial TISs highly localized near the interface. These states emerge in the middle of two mini-bandgaps with wavelengths of 567 nm and 581 nm, as detailed in the Supplementary Materials (SM). Besides, the SSD structure also allows for trivial defect states on both sides of the mini-bands due to the high refractive index region formed by three closely spaced ribbons at the interface. These trivial defect states are located on the defect polymer ribbon right at the interface [45].

The transmission polarization optical microscopy images of the fabricated SSD topological superlattice are shown in Fig. 1(b2, b3). Under the crossed polarizers, optically isotropic polymer ribbons always look dark. The CLC regions between the ribbons also look dark when the crossed polarizers are oriented parallel/perpendicular to the direction of the polymer ribbons, but they become bright when the two crossed polarizers are rotated by 45°. This indicates that the surface relief structures on the sidewalls of the ribbons enable good alignment of the CLC [58] and that the helical axis of the CLC is perpendicular to the direction of the ribbons. Our experimental setup for excitation and characterization of the topological lasing can be found in the SM. For inducing the lasing activity, the P-CLC superlattice is excited by a $Q$-switched frequency-doubled Nd-YAG laser operating at 532 nm with a repetition rate of 1.0 Hz and a pulse duration of 4.0 ns. The typical emission spectrum measured in our experiments is shown in Fig. 1(c), where one can see that there exist two lasing peaks at the wavelengths of 567 nm and 581 nm, corresponding to the two in-minigap TISs.

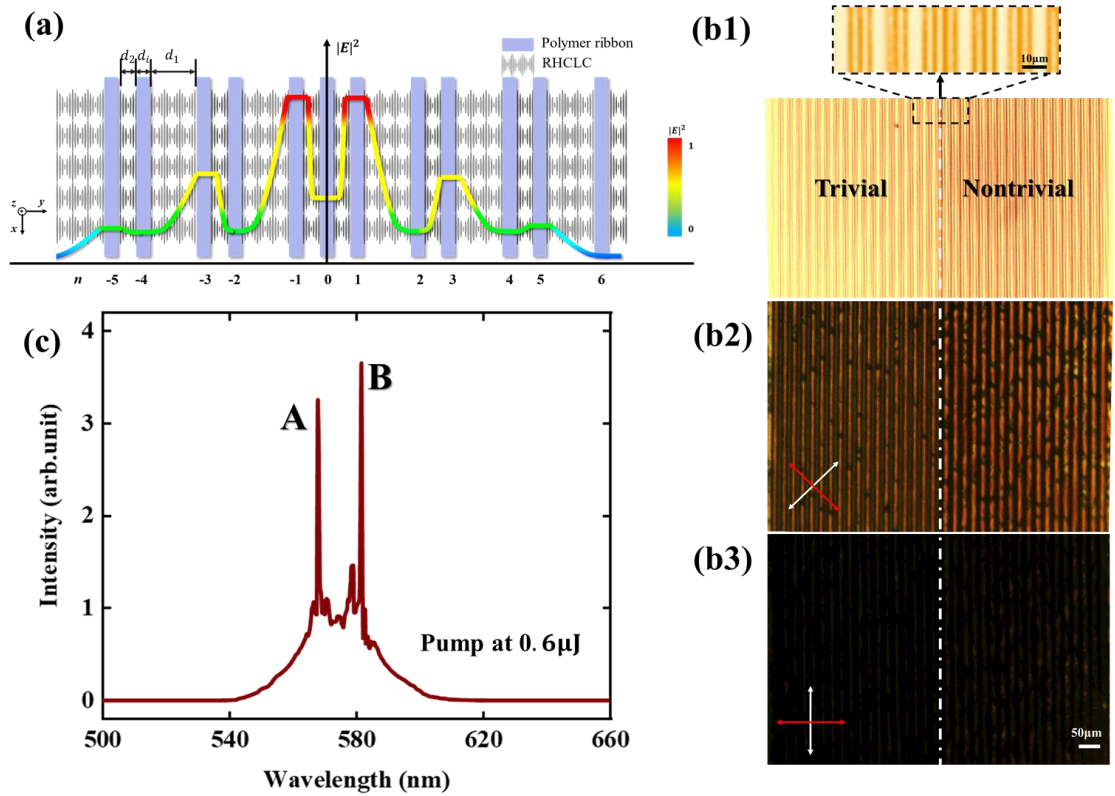

Fig. 1. **Fabricated topological P-CLC superlattice and interface-state lasing**. (a) A schematic diagram of the SSD topological structure and the intensity distribution of a TIS, where $n$ numbers the polymer ribbons. (b1) Standard optical microscopy image of the pure SSD polymer ribbon structure. The inset on the top shows an enlarged image around the interface. (b2, b3) Transmission polarization optical microscopy images of the P-CLC superlattice under two different cross-polarizer configurations. The white and red double-headed arrows indicate the orientations of the crossed polarizer and analyzer, respectively. (c) A typical TIS emission spectrum obtained from the superlattice at room temperature. A and B correspond to two TIS lasing peaks at 567 nm and 581 nm, respectively.

**Circular polarization and thermal tuning of topological emission**

Circularly polarized light with the same handedness as the helical structure propagating along the helical axis is selectively reflected by a CLC, while the opposite circularly polarized light is not affected by the structure. Hence selective photonic bands and bandgaps are formed only for circularly polarized light with the same handedness, so do the mini-bands and mini-bandgaps in P-CLC superlattices. The chiral characteristics of the constituent molecules and the corresponding helical arrangement of the RHCLC phase in this topological structure has the inherent property of inversion symmetry breaking, which provides a very favorable means for the realization of circularly polarized lasing. As shown in Fig. 2(a), the yellow-colored laser emission was collected at an angle of 90° to the incident direction of the pump light. To verify the circular polarization state of the topological lasing, we used a quarter-waveplate with its slow axis parallel to the vertical axis in combination with a linear polarizer to realize the polarization transformation. The obtained result is shown in the inset of Fig. 2(b). It reveals that the TIS lasing is transformed into linearly polarized light oriented at an angle of -45° to the horizontal axis, proving that right-handed circular polarization is generated in the system.

The CLC used in our experiments exhibits a cholesteric to isotropic transition (clearing point) at about 30°C. The elastic constants of the CLCs increase with decreasing temperature. Consequently, the torsional torque of the chiral dopant increases, resulting in the decrease of the CLC pitch. Therefore, decreasing temperature causes a blueshift of the optical bandgap. In addition, the width of the bandgap increases (see SM for more details). Figure 2(b) shows the thermal tuning of the lasing wavelength. When the temperature is decreased from 24°C to 8°C, topological lasing with wavelengths decreasing from 581 nm, 567 nm, 556 nm to 544 nm can be excited successively due to the blue shift of the bandgap. At lower temperatures, one can find some situations in which three TIS lasing peaks exist simultaneously. It is conceivable that, if required, single-mode topological lasing can be achieved by decreasing the bandgap width of the CLC and/or by increasing the mode spacing by using narrower polymer ribbons.

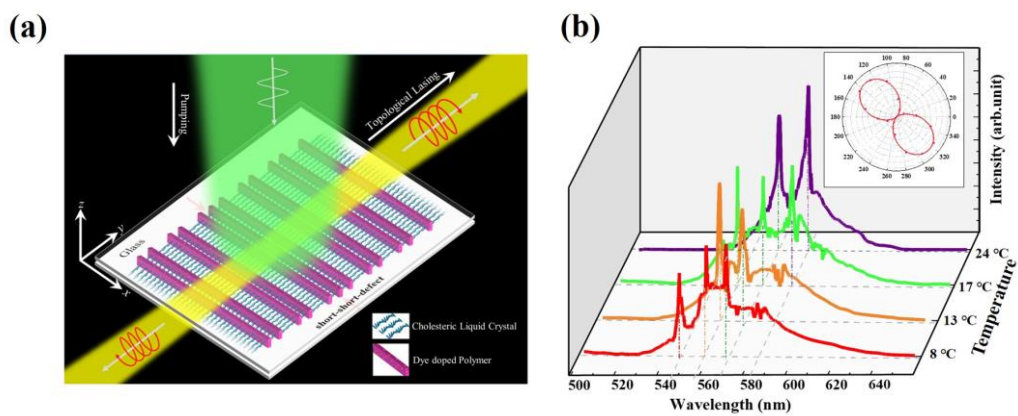

Fig. 2. **Illustration of TIS lasing in P-CLC assembly and thermal tuning.** (a) Schematic drawing of the TIS-lasing assembly, where the green pump beam is linearly polarized, and the yellow beams of laser emission in opposite directions are right circularly polarized. (b) Thermal tuning of the topological lasing wavelength. The inset shows the intensity of the quarter-waveplate-transformed laser radiation as a function of the polarization angle in the polar coordinate system. The polar angle stands for the transmission angle of the polarizer, and the radius stands for the transmittance.

**Robustness against perturbations**

We found that in our current fabrication method there exists about a 2% variation in the ribbon width and the spacing between the ribbons. To explore the effect of random fabrication errors on topological lasing, we excited the sample by steering the pump beam to five different interface positions. The resultant lasing spectra are shown in Fig. 3(a). They reveal that the TIS lasing always occurs and is preserved, although the lasing wavelength has a fluctuation within ±2.5 nm. This indicates that the TIS lasing is robust against the perturbations due to fabrication imperfections.

The TIS based on the SSH model is known to be robust against perturbations that respect chiral symmetry of the Hamiltonian [63-67]. The Hamiltonian of the SSH model containing an SSD by using the tight-binding method can be written as follow [47]:

$$H = v \left( \sum_{n \in N_+} (1 + \xi_{n+1}) a_n^\dagger a_{n+1} + \sum_{n \in N_-} (1 + \xi_{n-1}) a_n^\dagger a_{n-1} \right)$$

$$+ w \left( \sum_{n \in N_+} (1 + \xi_{n+2}) a_{n+1}^\dagger a_{n+2} + \sum_{n \in N_-} (1 + \xi_{n-2}) a_{n-1}^\dagger a_{n-2} \right) + h.c.$$

$$N_+ = 2N, \ N_- = -2N, \ N = (0, 1, 2, 3 \ldots) \tag{1}$$

where $a_n$ ($a_n^\dagger$) is the annihilation (creation) operator in the $n$-th site of the polymer ribbons labeled in Fig. 1(a), $\xi_n$ is the perturbation added to the coupling strength, $v$ and $w$ describe the strong and weak coupling coefficients between the polymer ribbons spaced by $d_2$ and $d_1$ respectively. The coupling strength can be effectively tuned by changing the spacing between neighboring ribbons, and a smaller spacing leads to a stronger coupling. In our SSD structure, $d_1 > d_2$ results in $w < v$. To verify the intrinsic robustness of TISs against perturbations, we add chiral perturbations ($\xi_{-n} = \xi_n$) on all off-diagonal terms of the Hamiltonian in Eq. (1). When $\xi_n = 0$, the coupling strength are $v$ = 0.0048 and $w$ = 0.0011 in our superlattices. To perform a quantitative analysis, we define the probability of the TIS (or the trivial defect state) coupled with the bulk states as $P = n_{bulk}/n_{all}$, where $n_{bulk}$ is the number of perturbation sets that result in coupling of the TIS (or the trivial defect state) with the bulk modes, and $n_{all}$ is the total number of perturbation sets. Then we respectively calculate the probabilities of the TIS and the trivial defect state coupling with the bulk states by adding $n_{all}$ = 500 sets of perturbations with $\xi$ equal to the maximum of $\xi_n$. Figure 3(b) shows the different behaviors of the TIS and trivial defect state under chiral perturbations in the SSD superlattice. One can see that even under a high perturbation strength ($\xi$ = 55%, nearly half ($\xi$ = 48%) of the trivial defect states couple with the bulk states, while all the TISs still remain independent and isolated from the bulk states. (see more details in SM). This result means that the TISs can still be localized even under high chiral perturbations ($\xi$ = 60%), which is beneficial for excellent topological lasing performance.

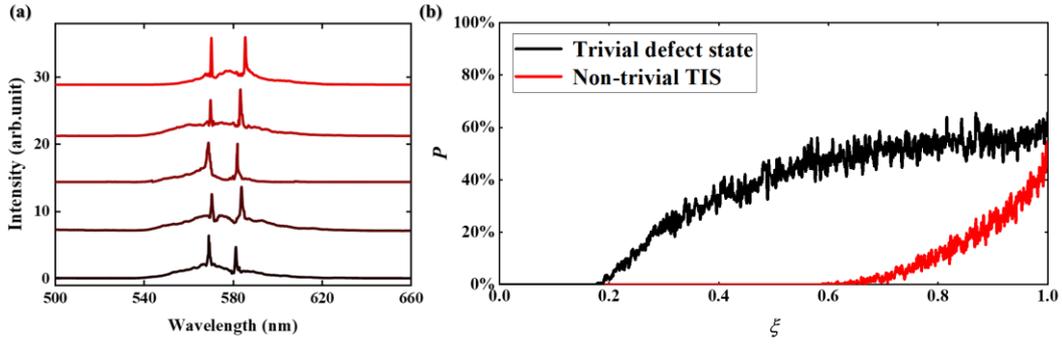

Fig. 3. **Influence of fabrication imperfection on TIS lasing and perturbation analysis**. (a) Measured emission spectra generated by optical pumping at different positions along the interface. (b) Plot of the coupling probability $P$ versus perturbation strength $\xi$, where red and black lines illustrate the TIS and the trivial defect state in the SSD structure, respectively.

**Comparison between TIS lasing and trivial defect state lasing**

Low lasing threshold is indispensable for practical applications. This has motivated tremendous efforts on exploration of defect-mode lasing based on CLCs, which was found to show a lower threshold compared to band-edge lasing of CLCs [68-71]. As we know, topological boundary (edge, corner, interface) states are more localized and robust than the conventional defect modes, which entails an even lower lasing threshold [17-33]. In addition, robust topological modes can maintain a high slope efficiency and high wavelength stability in the presence of defects and disorder [17, 38]. In our work, the highly localized TISs provide a favorable condition for realizing low threshold lasing. As can be seen from the illustration in Fig. 4(a), topological lasing appears with increased pumping energy. The lasing threshold observed in our experiments is about 0.4 μJ, corresponding to a peak intensity of 722 W/mm$^2$, which is almost four orders of magnitude lower than what has been achieved in previously reported visible topological laser [38]. According to Fig. 4(b), a clear decrease in the emission linewidth, i.e., spectral narrowing is observed with increasing pump energy, which further verifies the lasing behavior of the topological P-CLC superlattice.

Figure 4(c) shows the calculated photon density of states for the SSD superlattice, from which one can see that there exist peaks from the two TISs (A and B) and four trivial defect states (C, D, E and F). Both TISs and the trivial defect states localize around the interface of the SSD structure, although the modal distributions of the TISs are different from those of the trivial defect states [45, 46]. The densities of states of the TISs are higher than those of the trivial defect states, which empowers the TISs lower lasing thresholds. Thus, the independent excitation of TIS lasing under low pump energy (0.4 μJ - 1.18 μJ) is experimentally realized due to their lower thresholds with respect to trivial defect states. This is shown in the inset of Fig. 4(a), though TISs and trivial defect states are both present near the interface. As the pumping energy is further increased, the trivial defect states are also excited. As illustrated in Fig. 4(d), there exist two TIS lasing peaks with wavelengths at 567 nm and 581 nm, and four trivial-defect-state lasing peaks with wavelengths at 564 nm, 570 nm, 579 nm and 584 nm. These experimental results agree well with our theoretical analyses.

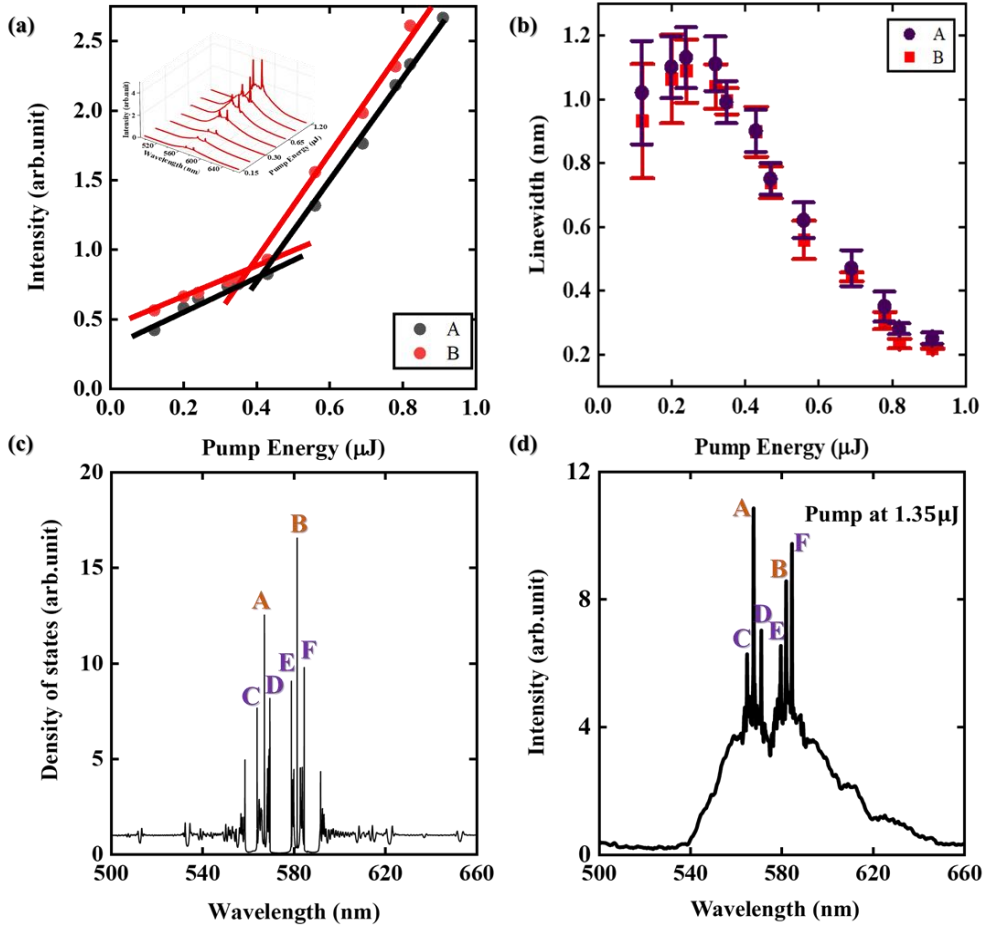

Fig. 4. **TIS lasing and trivial defect state lasing emission**. (a) The dependence of peak intensities of the TIS emission marked by A and B (see Fig. 1(c)) on the pump energy. (b) Linewidths of the peaks as a function of the pump energy. The error bar indicates a standard deviation of 5 measured spectra for each pump energy. (c) The calculated photon density of states for the SSD superlattice. (d) Emission spectrum obtained from the superlattice at room temperature pumped at 1.35 μJ. TISs are marked by A and B, and trivial defect states are marked by C, D, E and F.

## Discussion

We have demonstrated tunable topological lasing based on a P-CLC superlattice that generates circularly polarized radiation in the yellow spectral region. By splicing together two sublattices with different topological properties distinguished by the reversed mini-bands following the 1D SSH model, the TISs appear in these mini-bandgaps. Using the FLDW technology for polymer processing and the out-of-plane orientation technology for the CLC alignment, we have fabricated the predesigned SSD topological structure, dramatically reducing the fabrication difficulty and the costs.

Although the size of a primitive unit cell of polymer ribbons is at a micrometer scale, due to their combination with the CLC medium, visible topological interface state lasing has been experimentally realized, and the lasing wavelength is tuned by the temperature. Due to the high localization and robustness of the TISs, we have achieved a low threshold of the topological lasing at about 0.4 μJ (722 W/mm$^2$). We have showed that the lasing still exists even when there are relatively large perturbations present in the structure. Moreover, the coexistence of TIS lasing and trivial defect state lasing is observed and compared under a higher pump energy of 1.35 μJ. Such composite structures may play an important role for practical applications in the development of new tunable topological on-chip photonic devices, such as topological lasers, sensors, and quantum information processing units.

## Materials and methods

### Numerical simulation

The transmission spectrum and the band structure of the P-CLC superlattice are theoretically analyzed by using the adjusted Ambartsumian's layer addition modified method proposed by Vardanyan and Gevorgyan [72] as well as the Berreman 4×4 matrix calculation method [73]. For the calculation of electric field distribution, we use the transfer matrix method [74] and the simulation software based on the finite difference time domain (FDTD) method. The structural parameters of the simulations agree with experimental selection, but, to shorten the calculation time and improve the running speed, the number of structural periods considered in the simulation is smaller than that in the actual experiment. The CLC was a mixture of a commercial nematic liquid crystal (60 wt.%, Shijiazhuang Chengzhi Yonghua Display Material Co.) and a right-handed chiral agent CB15 (40 wt.%, Shijiazhuang Chengzhi Yonghua Display Material Co.) with a clearing point at about 30°C. The refractive indices of the CLC used in the calculations were 1.7 and 1.6 for extraordinary and ordinary light, respectively, and the refractive index of the polymer ribbons was taken to be 1.55.

### Fabrication

A negative photoresist SU-8 (model 3010, MicroChem) doped with a laser dye PM597 (0.2wt.%, Pyrromethene 597, Exciton Inc.) was spin-coated on an ITO-coated glass plate. A titanium-doped sapphire ultrashort pulsed laser with a central wavelength of 800 nm, a pulse width of 80 fs, and a repetition frequency of 80 MHz was used to fabricate the polymer ribbon structures. According to the morphology characterization of the ribbons, the thickness of a single ribbon is about 2 μm, and the height is about 10 μm. After the preparation of polymer ribbons, another ITO-coated glass plate with a PDMS overlayer was used to cover the polymer scaffold, and the assembly was glued together at the edges. Then the CLC was filled into the space between the ribbons by a capillary force, and the entire "CLC cell" was sealed from all sides with a UV curable adhesive. Due to the diffraction effects during the FLDW process, the minimum spacing between the polymer ribbons is limited to about 4.0 μm, in order to obtain appropriate channels for good CLCs filling. Due to this limitation, in the dimer and the

SSD structures, the distance between weakly coupled ribbons was chosen as $d_1 = 34 \cdot (p/2)$, while that between strongly coupled ribbons was $d_2 = 22 \cdot (p/2)$.

**Acknowledgments**
This work is supported by the National Key Research and Development Program of China (2017YFA0303800, 2020YFB1805800), National Natural Science Foundation of China (12134006, 12074201, 12104335), 111 Project (B07013), PCSIRT (IRT_13R29), and Slovenian Research Agency research program P1-0192.

**Author contributions**
X. Z. Z., Z. G. C. and J. J. X. conceived and supervised the project. Y.W. and S. H. G. were responsible for the numerical simulations. Y. W. and D.H.Y. fabricated and characterized the samples. I. D., Q.W. and M. C. participated in the measurements and discussions. Y.W. wrote the paper with contributions from all authors. All authors reviewed the manuscript.

**Conflict of interest**
The authors declare no competing interests.